\documentclass[aps,prl,twocolumn,superscriptaddress]{revtex4}

\usepackage{graphicx}  
\usepackage{dcolumn}   
\usepackage{bm}        
\usepackage{amssymb}   
\usepackage{verbatim}  
\usepackage{color}
\usepackage[gen]{eurosym}

\begin{document}

\title{Another approach to the study of phase noise
in electrical oscillators}

\author{Marcello Carl\`a}
\email{carla@fi.infn.it}
\affiliation{Department of Physics, University of Florence,
Via G. Sansone 1 50019 Sesto Fiorentino - (FI) - Italy}

\date{\today}

\begin{abstract}

The mechanism at the base of phase noise generation in electrical
oscillators is reexamined from first principles. The well known
Lorentzian spectral power distribution is obtained, together with
a clearcut expression for the line-width parameter.
The mechanism of suppression of the amplitude fluctuations and
its effects are discussed and
the true role of the figure of merit $Q_0$ of the resonator is
restated.
The up conversion of low frequency components from a non-white
noise source is also considered.
A number of simple numerical experiments
is presented to illustrate and clarify the mathematical results.
\end{abstract}

\maketitle

\section{introduction}

Signal generation is half the world of electronics,
the electronic oscillator is ubiquitous, it has a part in all consumer,
professional and scientific appliances and most often the quality of
the generated signal is crucial to the overall performances.

After the problems of accuracy, stability, adjustability and resolution,
one of the main concern in today signal generation is spectral purity,
either harmonic content or non-harmonic wide and narrow band noise.

The fact that a sinusoidal signal at the output of an oscillator
never is truly \emph{monochromatic} in the mathematical
sense, but has a \emph{line-width} and noise side bands, has important
implications.
For example, it reduces the possibility for a radio receiver
to discriminate a week signal close to a strong one (\cite{2012-Pozar}).

The first paper on this subject I am aware of is chapter 15 in Edson's
``Vacuum Tube Oscillators'' \cite{1953-Edson}.
In this work Edson assumed that an electronic oscillator could
be considered an almost linear
system whose nonlinearities gave a negligible contribution to the
spectral composition of the signal.

Edson's model for an oscillator was a sort of enhanced Q-multiplier
brought to
the limit of hugely amplifying and filtering the thermal noise always
present in any real circuit.
In this model the gain of the positive feedback loop was smaller than unity
by an infinitesimal amount, to accomodate for the noise power at
the amplifier input.
The equivalent $Q$ of the circuit resulted very very high, but
not infinite, and its selectivity accounted for the Lorentzian
shape of the generated signal spectrum, in agreement with
experimental observations.

This model could account for the presence of the narrow noise
skirt on sides of the generated signal (the carrier) and for the 20 dB
intensity decay per decade of frequency moving away from the
carrier.

This work did receive little attention and was little quoted in the
literature in the years following.
The author himself declared that he ``was never fully satisfied with
the results from this approach ...'' \cite{1983-Edson},
to the point that, a few years after the book, he published a
completely new paper on the same argument \cite{1960-Edson}.

Moreover, experimental studies soon demonstrated the presence of a
$1/f$ noise
component in the frequency region nearest to the carrier, that Edson's
model could not explain, not even including some amount of non
linearity.
Yet, Edson's model had the great virtue of being very clear to
understand and of directly connecting
noise characteristics with electrical circuit parameters.

Other works followed shortly after.
In \cite{1966-Leeson} Leeson approached the problem from an heuristic
point of view, presenting an equation that described the commonly
observed spectrum of feedback oscillators.
This paper, though very popular in the following works, did not face
the problem of connecting the observed noise spectrum properties with
the electrical circuit characteristics.
Little later Lax laid the basis for approaching the problem
through statistics and interest into the argument
was paralleled by an analogous interest for the line width of a
laser generated light beam \cite{1967-Lax}.

Recently interest into oscillator noise has greatly increased because of
its relevance in the performance of telecommunication networks, both
digital and analog, wired and wireless, as documented by the
large number of recent contributions to the literature. See, e.g.,
\cite{1998-Hajimiri,2000-Lee,2003-Ham,
2010-Mirzaei} and references therein.

These works, as well as Lax's one, greatly deepened
the theoretical analysis
of the stochastic processes behind the phase noise origin.
They make large use of concepts and arguments from the world of
statistics, at the expense,
though, of somewhat  missing a clearcut connection with the physical
circuit, its parameters and its behaviour.

In \cite{2003-Ham}, e.g., the analogy
has been used between the oscillator signal
phase fluctuations and the process of molecular diffusion.
As suggestive and appropriate as it can be by principle such
a connection, its application to the study of phase noise
in oscillators can help only from a qualitative point of view.
Otherwise, a double effort would be required first to correlate
circuit equations with molecular processes and then to bring results
from molecular behaviour back into the circuits realm.

A consequence of this state of affairs from a didactic point of view
is that the subject is often relegated to books and courses of electronics
at the highest specialization level or is proposed through
an heuristic and qualitative only approach, following Leeson's approach
(see, e.g., \cite{2012-Pozar}).

In this paper I will try to approach the problem making profit of
the large and good amount of work already available, through
a simple but rigorous enough way, making use mostly of those concepts
familiar to the world of electronics like
Johnson's noise, modulation,
frequency conversion and vectorial (de)composition.
Clearly, some mathematical extras to this menu are unavoidable,
not to remain at the qualitative and descriptive level only.
They are the Central Limit and the Wiener-Khinchin theorems and
some properties of gaussian distributions.
Luckily enough, currently
they are well established and widely known theorems, also at the
student level.

The path I shall propose in this paper
easily lends itself to the implementation
of numerical experiments, very useful from a
didactic point of view to illustrate the theoretical concepts
and enhance confidence into the results obtained through equations.

\section{The model oscillator}

It is a common practice in a linear system to group all noise sources
into a single one, conveniently placed and with an equivalent effect on
the output signal.

In this paper I have chosen not to follow this way for the sake of
the simmetry both in the circuit and in the equations that will be written.
The cost of this choice will be that most equations will contain
the sum of two equivalent terms, one for current, one for voltage,
with no further increase in complexity.
The advantage will be a clearer splitting of noise into amplitude
and phase components and the possibility to introduce into
the circuit some non linear behaviour without having to deal with the
violation of the primary hypothesis ``in a linear system''
intrinsic to the common practice.
In the last, as expected, it will be clear that the total noise power is
the unique meaningful quantity in evaluating phase noise.

In an electronic oscillator, like in every harmonic oscillator, unavoidable
sources of noise are the dissipative elements associated with the two
energy storing components of the resonator, most
often an inductance $L$ and its series
resistance $R$ and a capacitance $C$ and its parallel conductance $G$
(fig.\ 1).

The dissipative elements $R$ and $G$ are responsible both for
the introduction of thermal noise into the circuit,
a random voltage $v_n(t)$ and current $i_n(t)$ associated with
resistance $R$ and conductance $G$, and for the resonator damping.

\begin{figure}[!htb]
\includegraphics[width=0.8\columnwidth]{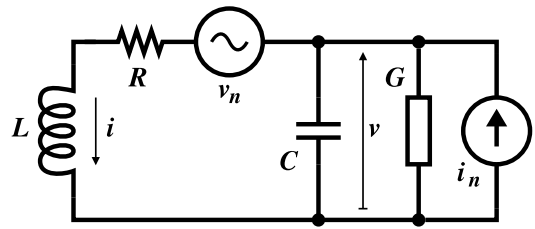}
\caption{
\label{fig-1}
{A typical resonator in an electronic oscillator is composed by
an inductance $L$ and a capacitance $C$, together with their
resistance $R$ and conductance $G$, that account for losses.
The two generators $i_n$ and $v_n$ are the noise sources inseparably
linked to the dissipative elements.}
}
\end{figure}

Actually, in a stationary oscillator values of damping and noise
generation are disjoined from the physical values of $R$ and $G$,
because of circuits and devices that provide the positive
(regenerative) feedback indispensable to add
a negative damp that cancels out the natural (positive) damp of the
resonator and allows a persistent oscillation to be sustained.

These circuits and devices introduce their own noise contributions
that add to $i_n$ and/or $v_n$.

As a consequence, the value of $R$ and $v_n$ and of $G$ and $i_n$
can be considered uncorrelated, only with the constraint  that
the spectral density $V_n(\omega)$ and $I_n(\omega)$ of
$v_n(t)$ and $i_n(t)$ cannot be lower than what due respectively to the
$R$ and $G$ of the physical circuit, namely
\[
V_n \ge \sqrt{4 k T R} ~~~~~~~~
I_n \ge  \sqrt{4 k T G}
\]
where $k$ is Boltzmann's constant and $T$ absolute temperature.

So, it is possible to consider the practical case where
damping is kept to zero by regeneration,
i.e. $R$ and $G$ are nulled by equivalent negative terms,
but some noise, both $v_n$ and $i_n$, is present in the circuit.
After putting some energy $W_0$ into $L$ and $C$ there will be a
persistent stationary oscillation.
For the sake of simplicity, let's consider the initial conditions
$v(t)=0 \ V$ and $i(t)=1 \ A$ at $t=0$ so that, in the absence of noise,
oscillation would be
\begin{eqnarray}
v(t) & = & v_0 \sin(\omega_0 t) \nonumber \\
i(t) & = & i_0 \cos(\omega_0 t) \nonumber
\end{eqnarray}
with, as usual, $\omega_0 = 2 \pi f_0 = 1 / \sqrt{LC}$ and
\[
{1 \over 2} L i^2(t) + {1 \over 2} C v^2(t) =
{1 \over 2} L i_0^2 = {1 \over 2} C v_0^2 = W_0
\]

In the state space described by the $i,v$ circuit variables
in absence of noise the representative
point $S$ of the system describes, as well known,
a circle with constant angular velocity $\omega_0$.
In the presence of noise both the trajectory and angular velocity of the
point are perturbed, as described in fig.\ 2.

\begin{figure}[!htb]
\includegraphics[width=0.8\columnwidth]{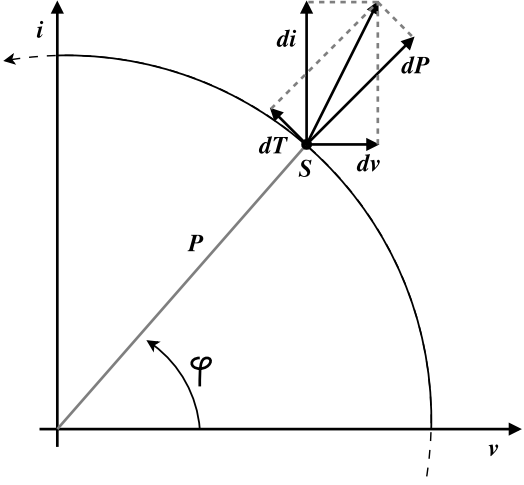}
\caption{
\label{fig-2}
Decomposition of independent noise contributions $dv = (i_n/C) dt$ and
$di = (v_n / L)dt$ into the radial $dP$ and tangential $dT$ parts.
Normalization of the $i$, $v$, $di$ and $dv$ variables over $i_0$ and $v_0$
as expressed by eq.\ (\ref{eq:Psq})
is not shown to avoid cluttering the figure
and should be implicitly assumed.}
\end{figure}

\subsection{Noise decomposition}

By effect of noise current $i_n$ and voltage $v_n$,
in a time $dt$ charge upon capacitor $C$
and current through inductor $L$
change by an amount (with effective $G=0$ and $R=0$)
\begin{equation}
dv = {i_n \over C}\ dt
~~~~~~~~~~~~
 di = {v_n \over L}\ dt
\label{eq:dv-di}
\end{equation}

As $v_n$ and $i_n$ are statistically independent random variables, so are
$dv$ and $di$.

Whatever their instantaneous values, $dv$ and $di$ can always be decomposed
into a
tangential $dT$ and a radial $dP$ part as shown in fig.\ 2, with:
\begin{equation}
P^2 = \left( i \over i_0 \right)^2 + \left( v \over v_0 \right)^2 =
{{{i^2 L} + {v^2 C}} \over {2 W_0}}
\label{eq:Psq}
\end{equation}

The effects of $dT$ and $dP$ over the motion of point $S$ are
independent each other: $dP$ changes the oscillation
amplitude $P$, $dT$ its phase $\varphi$ by the amount $d\varphi = dT / P$.

There is only a small link left between phase and amplitude fluctuations.
If the noise generators $i_n$ and $v_n$ are stationary, as usual, their
\emph{rms} amplitudes remain constant.
Hence, the same amount of noise from $i_n$ and $v_n$ yields a different
effect on $d\varphi = dT / P$, depending upon the instantaneous value of
amplitude $P$.
But this effect is vanishingly small indeed.
In a real system the standard deviation
of the $P$ fluctuations is many orders of magnitude smaller than the $P$
average value: nobody would be concerned in a change of one part per million
of a noise source relative intensity.

By standard analytical geometry relations, the transformation
\begin{eqnarray}
dP & = &
\left( \phantom{-} {i_n \over {v_0 C P}} \cdot \cos{\varphi} +
{v_n \over {i_0 L P}} \cdot \sin{\varphi} \right) dt \label{eq:dP} \\
dT & = & \left( - {i_n \over {v_0 C P}} \cdot \sin{\varphi} +
{v_n \over {i_0 L P}} \cdot \cos{\varphi} \right) dt
\label{eq:dT}
\end{eqnarray}
holds. Thereafter, by transforming a couple of values $(v_n, i_n)$ by
eqs.\ (\ref{eq:dP}-\ref{eq:dT}), zeroing
alternatively the $dP$ or $dT$ part and transforming back with
the inverse of (\ref{eq:dP}-\ref{eq:dT}),
two couples of components $(v_{nt}, i_{nt})$ and
$(v_{np}, i_{np})$ are obtained that sum up to give
the original $(v_n, i_n)$ values,
but yield each one only a $dT$ or $dP$ perturbation:
\begin{eqnarray}
v_{nt} & = & - i_n R_0 \sin{\varphi} \cos{\varphi} + v_n \cos^2{\varphi}
\label{eq:vnt} \\
i_{nt} & = & i_n \sin^2{\varphi} - {v_n \over R_0} \sin{\varphi} \cos{\varphi}
\label{eq:int} \\
\nonumber \\
v_{np} & = & i_n R_0 \sin{\varphi} \cos{\varphi} + v_n \sin^2{\varphi}
\label{eq:vnp} \\
i_{np} & = & i_n \cos^2{\varphi} + {v_n \over R_0} \sin{\varphi} \cos{\varphi}
\label{eq:inp}
\end{eqnarray}
with $R_0 = \sqrt{L/C}$.

Using these equations it is possible to decouple the $dT$ and $dP$
components of the random
motion of the representative point $S$ in the state space
and go on studying their effect independently each other.

Clearly, in a completely linear circuit as in fig.\ 1,
such a decomposition is only a mathematical formalism.
Actual voltage and current will be given only by the sum of the
two components and will be the linear response of the
\emph{LRCG} network under the action of the $v_n$, $i_n$ generators.

But, when the circuit will be modified in order to suppress
one of the two noise components, namely the amplitude fluctuations,
and this can be obtained only through the
unavoidable introduction of a suitable non linear behaviour,
then the remaining tangential component (the phase fluctuations)
will describe the true physical current and voltage.

\subsection{Phase fluctuations}

By definition of angular frequency and using eq.\ (\ref{eq:dT})
\begin{equation}
{{d \varphi}} = \omega_0 \cdot dt + 
\left( - {i_n \over {v_0 C P}} \cdot \sin{\varphi} +
{v_n \over {i_0 L P}} \cdot \cos{\varphi} \right) \cdot {dt \over P}
\label{eq:d_phi}
\end{equation}
This equation expresses how noise voltage and current from the
$v_n$ and $i_n$ sources  induce fluctuations in the phase $\varphi$
of the oscillator signal.
The two quantities $v_0 C P$ and $i_0 L P$ represent the amplitude of the
oscillation (the envelope) and its evolution over time.
In any real oscillator of practical interest, the amplitude is controlled
by some kind of stabilization mechanism, as described in the next
section;
hence in eq.\ (\ref{eq:d_phi})
amplitude drift can be neglected and $P$ does not move from the
initial value $P = 1$.
This yields
\begin{equation}
\omega = {{d \varphi} \over {d t}} = \omega_0 + 
\left( - {i_n \over {v_0 C}} \cdot \sin{\varphi} +
{v_n \over {i_0 L}} \cdot \cos{\varphi} \right)
\label{eq:omega}
\end{equation}
The effect of the phase fluctuations is clearly the introduction of a
frequency modulation over $\omega_0$.
In the conditions that are common in real (and useful) oscillators
this effect is very very small indeed, hence it is
safe to write
\begin{equation}
\omega_n = \omega - \omega_0 =
 - {i_n \over {v_0 C}} \cdot \sin(\omega_0 t) +
{v_n \over {i_0 L}} \cdot \cos(\omega_0 t)
\label{eq:omega-n}
\end{equation}
and
\begin{equation}
\varphi(t) = \omega_0 t + \varphi_n(t)
~~~~~~~~~
\varphi_n(t) = \int_0^t{\omega_n(t') dt'}
\label{eq:phi(t)}
\end{equation}
(but in the numerical experiments the differential equations of the circuit
will be integrated without using such an approximation).

If $i_n$ and $v_n$ are statistically independent sources of white noise,
as it is the case for Johnson's noise in resistors, they have a uniform
spectral density.
Hence, the product of their components with the $\sin()$ and $\cos()$
oscillations at frequency $\omega_0$ has the effect of folding their
constant spectral density over itself, yielding again a constant
spectral density.
Under such conditions, the spectral density as a function of frequency
$f$ of the resulting oscillation around frequency
$f_0$ is the Lorentzian curve
(normalized to unitary power over the full frequency range)
\begin{equation}
\mathcal{L}(f) = {1 \over {\pi \sigma}} \cdot {1 \over
{\displaystyle \left( {f - f_0} \over \sigma \right)^2 + 1}}
\label{eq:lorentzian}
\end{equation}
where $\sigma$ is the \emph{line-width} of the generated signal.

The path from eq.\ (\ref{eq:omega-n}) to (\ref{eq:lorentzian})
contains all the mathematical difficulties, so
widely described in the literature.
It is possible to travel this path in three steps, making use of two
widely known theorems (Central Limit and Wiener-Khinchin theorem)
and an established property of gaussian distributions.
All of this will be used here without any demonstration, that is
left to appropriate textbooks.

The starting point is the requirement that $i_n$ and $v_n$ be white
noise sources.
It is not necessary they have a Gaussian distribution.
Usually, $\omega_n$ will not have a Gaussian distribution, apart
some very special conditions, because the product of $i_n$ and $v_n$
with the $\sin()$ and $\cos()$ functions in eq.\ (\ref{eq:omega-n})
would alter their original Gaussian distribution, if any.
In any case, $\varphi_n$ \emph{will have} a Gaussian distribution, as
stated by the Central Limit theorem, because $\varphi_n$
results from the sum
of infinitely many independent random contributions, as seen
in eq.\ (\ref{eq:phi(t)}).

The second step is to obtain the time autocorrelation function
$R_F(\tau)$ for the signal at the output of the oscillator,
namely for the $\sin[\varphi(t)]$ or the $\cos[\varphi(t)]$ function.

Following \cite{2010-DiDomenico,1982-Elliott},
using the complex notation, this can be written as
\begin{eqnarray}
R_F(\tau) & = & {A_0^2 \over 2} \cdot \mathrm{Re}
\left\{ \lim_{T \to \infty}
{1 \over T}\int_{- {T \over 2}}^{+{T \over 2}}
e^{i \varphi(t)} \cdot e^{-i \varphi(t + \tau)} dt \right\} \nonumber \\
& = &  {A_0^2 \over 2} \cdot \mathrm{Re} \left\{ e^{-i \omega_0 \tau}
\left<
e^{i \left[ \varphi_n(t) - \varphi_n(t + \tau) \right]}
\right>
\right\} \nonumber
\end{eqnarray}
where $A_0$ is the oscillation amplitude and the $\left< \right>$
brackets are a short for the $\left\{ \lim\int \right\}$ computation.

Here the properties of the gaussian distribution play their part.
Using the gaussian moment theorem, it is possible to write
\begin{equation}
\left<
e^{i \left[ \varphi_n(t) - \varphi_n(t + \tau) \right]}
\right> =
e^{- {1 \over 2}
\left< \left[
\varphi_n(t) - \varphi_n(t + \tau)
\right]^2 \right>
\label{eq:m_theorem}
}
\end{equation}

A short but clear demonstration for this can be found in \cite{1982-Elliott}
and references therein.

Evaluation of the exponent in the right hand side of
eq.\ (\ref{eq:m_theorem}) is easier if it is made in a discrete time
framework, like in the numerical integration of the circuit
differential equations in the numerical experiments.
It has been supposed that noise generators are stationary, hence
time average can be replaced by variance (indicated by the overline):
\begin{equation}
\left< \left[
\varphi_n(t) - \varphi_n(t + \tau)
\right]^2 \right> =
\overline{ \left[ \varphi_n(0) - \varphi_n(\tau) \right]^2 } =
\overline{\varphi_n^2(\tau)}
\end{equation}

At every $\Delta t$ time step, $\varphi_n$ increases by the amount
$\omega_n \Delta t$.
Increments are random and uncorrelated, hence sum quadratically.
After $N = \tau / \Delta t$ time steps
\begin{equation}
\overline{\varphi_n^2(\tau)} = \overline{\omega_n^2} \Delta t^2
{\tau \over {\Delta t}} =
\overline{\omega_n^2} \Delta t \ \tau
\label{eq:phi-2n}
\end{equation}

It is to be recognized that $2 \overline{\omega_n^2} \Delta t$
is the spectral density of $\omega_n$
(this will be made clear in the description of the numerical experiments):
\begin{eqnarray}
\label{eq:Omega-n2}
2 \overline{\omega_n^2} \Delta t =
\Omega_n^2 & = & {1 \over 2} \left(
{I^2_n \over {v_0^2 C^2}} + {V^2_n \over {i_0^2 L^2}}
\right) \\
& = &
{1 \over {4 W_0}} \left(
{I^2_n \over C} + {V^2_n \over L}
\right) \nonumber
\end{eqnarray}
where $\Omega_n^2$ has dimensions of $s^{-1}$.

The final form of the autocorrelation function is
\begin{equation}
R_F(\tau) = {A_0^2 \over 2} \cos (\omega_0 \tau)
e^{- \Omega_n^2 \tau / 4}
\label{eq:S(tau)}
\end{equation}

The third and last step is the application of Wiener-Khinchin theorem
that states that the spectral distribution $F(\omega)$ of a stationary random
sequence is linked to the Fourier transform of its autocorrelation
function:
\begin{equation}
F(\omega) = 4 \int_0^\infty R_F(\tau) \cos(\omega \tau) d\tau
\end{equation}
Applying the theorem to function (\ref{eq:S(tau)})
\begin{eqnarray}
F(\omega) & = & 2 A_0^2 \int_0^\infty \cos(\omega \tau) \cos(\omega_0 \tau)
e^{-\Omega_n^2 \tau / 4} \nonumber \\
& = &  A_0^2 \int_0^\infty e^{-\Omega_n^2 \tau / 4} \cdot \\
& & \left\{ \cos[(\omega_0 - \omega) \tau] +
     \cos[(\omega_0 + \omega) \tau] \right\} d\tau \nonumber
\end{eqnarray}
The rapidly oscillating term with frequency $\omega_0 + \omega$
gives a negligible contribution to the integral and can be dropped
(the integral goes rapidly to zero when $\omega$ moves away from
$\omega_0$).
Developing the integral of the remaining term,
the Lorentzian of eq.\ (\ref{eq:lorentzian}) is obtained, with
\begin{equation}
\sigma =  {\Omega_n^2 \over {8 \pi}}
\label {eq:sigma}
\end{equation}

This result is illustrated in the Numerical Experiments paragraph,
in the first numerical experiment, where only the components
($v_{nt}$, $i_{nt}$) from the full noise sources ($v_n$, $i_n$) will be applied
to the resonant circuit.
The oscillation amplitude will remain constant and the well known
Lorentzian noise distribution will appear on sides of $\omega_0$.

\subsection{Amplitude fluctuations}

As anticipated above,
every oscillator of practical interest has some kind of regulation
mechanism that controls the oscillation amplitude.

There is a twofold requirement for this.
First, the condition for persistent oscillation is that the damping given
by $R$ and $G$ be canceled out (it does not make a great difference
if each one separately, as considered in this work, or together as a
whole).
This is accomplished, as well known, using a positive feedback,
that introduces into the circuit a negative
$R'$ and/or $G'$ that \emph{exactly} compensates the $R$ and $G$
damping effect.

The word \emph{exactly} is a term out of the realm of physics or
engineering; however accurate a compensation is made, it will not be
\emph{exact} but for a short instant, thereafter a drift of one
kind or another will make damping move away from zero and make the
oscillation rise or decay.

Only a negative feedback regulation mechanism can continuously
compare actual oscillation amplitude with a reference value and
adjust the negative damping to contrast any drift.

Indeed, in a matematical sense damping can be made exactly null.
This has been assumed in all the numerical experiments,
not inserting into the oscillator circuit either $R$ or $G$.
But in the second numerical experiment the two noise sources
$i_n$ and $v_n$ will exchange
energy with the $LC$ resonator and make the oscillation amplitude
fluctuate in a random way.

The current generator $i_n$, for example, will induce a disturbance in
the voltage across condenser $C$ whose spectral amplitude will be
\begin{equation}
V_{cn}(\omega) = I_n (\omega) \cdot {1 \over
\displaystyle \omega C - {1 \over {\omega L}}}
\label{eq:vcn}
\end{equation}

At the resonance frequency $\omega_0 = 1 / \sqrt{LC}$ expression
(\ref{eq:vcn})
diverges, and indeed this is what would happen, given an infinitely
long time.

But the presence of the regulation mechanism, that cannot be avoided
for the reasons explained above, keeps under control and limits the
amplitude fluctuations also.
And this is the second reason the regulation mechanism is needed.

One of the simplest conceivable regulation mechanism is a first order
control loop described by the equation
\begin{equation}
{dP \over dt} =  K \cdot (P_0 -P)
\label{eq:dPdt}
\end{equation}
that has a stationary asynthotic solution $P = P_0$ and a transient
behaviour given by an exponential relaxation with time constant
$1 / K$.

The dynamical behaviour of the solution of
this equation is analogous to the oscillation amplitude
decay in a resonant
circuit with resonance frequency $\omega_0$ and figure of merit $Q_0$:
\[
{d \over dt} v_0(t) + {\omega_0 \over { 2Q_0}} v_0(t) = 0
\]
with
\begin{equation}
K = {\omega_0 \over {2 Q_0}}
\label{eq:K}
\end{equation}
Similar equations hold for $i_0(t)$.

Hence the behaviour we should expect for the noise induced amplitude
fluctuations is the excitation of a resonance around frequency $\omega_0$
with a bandpass $\pm \Delta \omega = K$.
This process is akin to the \emph{virtual damping} in \cite{2003-Ham} and
will be illustrated in the second numerical experiment.

The scenario outlined up to now is completely consistent if it is
recognized that considering only the ($v_{nt}$, $i_{nt}$) couple of
components of the ($v_n$, $i_n$) noise in the circuit while
discarding the ($v_{np}$, $i_{np}$) components is fully equivalent
to suppressing such fluctuations through a very efficient amplitude
regulation mechanism, as expressed by eq.\ (\ref{eq:dPdt}).

\subsection{Upconversion of $1/f$ noise}

From eqs.\ (\ref{eq:omega}-\ref{eq:omega-n})
it is clear that if $i_n$ and $v_n$ are \emph{coloured}
noise sources, also the resulting spectrum for $\omega_n$ will
be coloured, hence the main mathematical assumption in derivation
of result (\ref{eq:lorentzian})
will no more hold and the spectrum of the generated
signal mostly will not be a Lorentzian curve.

It is out of the purpose of this paper to analyze in any detail the
implications of eqs.\ (\ref{eq:omega}-\ref{eq:omega-n})
with non-white $v_n$ and $i_n$ noise sources.
For a deeper analysis see, e.g., \cite{2010-Mirzaei}.
Only an attempt has been made to illustrate the point, with the third
numerical experiment where coloured noise sources have been used for
$v_n$ and $i_n$.

\section{Numerical experiments}

Numerical experiments have been carried out using the circuit of
fig.\ 1 and numerically integrating the circuit differential equations
in the presence of the two noise generators $i_n$ and $v_n$.
The computational engine of the Spice program \cite{Spice} is
a very well optimized computer application to perform this task.
In this work all computations have been made using version 24 of
\emph{ngspice} \cite{ngspice}, an actively maintained open source variety
of Spice, and the
\emph{fftw} library \cite{fftw} to compute by Fast Fourier Transform (FFT)
frequency spectra from temporal sequences (this library also
available as open source).

All computations that will be described can be afforded using
a medium-high range PC, but their repetition to improve results
quality by averaging noise spectra can be very time consuming.
Hence, most of the repetitive work has been performed using
an HP ProLiant DL560 Gen8 machine equipped with 64 processors
and $128 \ GB$ RAM.

\bigskip

The basic machinery used for the first numerical experiment
can be seen in the following Spice list:

\begin{verbatim}
*** experiment 1 ***

l1   0  2   159.154943092e-6 ic=1
c1   1  0   159.154943092e-6 ic=0
vnv  3  0   dc 0 trnoise(100m 10u 0 0 0 0 0)
vni  4  0   dc 0 trnoise(100m 10u 0 0 0 0 0)
bwn  5  0   v=v(1)*v(3)-i(bvg)*v(4)
bvg  2  1   v=v(1)*v(5)
big  1  0   i=-i(bvg)*v(5)

.control
set nobreak
set numdgt=12
tran 10u  10000m  0m  0.1u  uic
linearize
print v(1) i(l1) > data.out
quit
.endc
\end{verbatim}

Values for $L$ and $C$ (\verb+l1+ and \verb+c1+)
have been chosen to give a resonance frequency $f_0$
as close as possible to $1000 \ Hz$ to reduce the noise generated by
truncation during FFT (in the experiments when phase is not affected)
and have been set numerically equal to
avoid unneeded computational overheads.
This way equal amplitudes on \verb+vnv+ and \verb+vni+ give
equal contribution to overall noise power.
There should be no loss of generality in this choice.

The two generators \verb+vnv+ and \verb+vni+ are white noise voltage
sources that yield random sequences of values with a gaussian
amplitude distribution and \emph{rms} value $V_G = 0.1 \ V$.
The cadence of the computed signals has been set to
$\Delta t = 10 \mu s$, hence Nyquist's frequency is
$f_N = 1 / (2 \Delta t) = 50 \ kHz$ and
the resulting noise spectral density is
$V_G^2 / f_N = 2 V_G^2 \Delta t = 0.2 \cdot 10^{-6} \ V^2_{rms} / Hz$
($-67 \ dBV$ in a $1 \ Hz$ bandwidth).

In a similar way the random variable $\omega_n$ of eq.\ \ref{eq:omega-n}
with variance $\overline{\omega_n^2}$ has spectral density
$\Omega^2_n = 2 \overline{\omega_n^2} \Delta t$ as given in
eq.\ \ref{eq:Omega-n2}.

Generator \verb+bwn+ is a convenience to obtain an intermediate
result and can be used to draw distributions of $\omega_n$,
as in fig.\ 5 below.

Two further controlled generators (\verb+bvg+ and \verb+big+) transform
the independent voltages from \verb+vnv+ and \verb+vni+
into the pair of values $v_{nt}$, $i_{nt}$
given by equations (\ref{eq:vnt}-\ref{eq:int})
and play the role of $v_n$ and $i_n$ in the circuit.

Because the numerical values of $L$ and $C$ have been chosen
to be equal, if the amplitude of the oscillation
remains constant and equal to $1$ (both $V$ and $A$),
with the given initial conditions
$\sin{\varphi}$ and $\cos{\varphi}$ can be written simply as
\[
\sin{\varphi} = i     ~~~~~~~~~  \cos{\varphi} = v
\]
namely, $\sin{\varphi} = \verb+i(bvg)+$ and
$\cos{\varphi} = \verb+v(1)+$.

First of all, the computational engine has been tested \emph{idle},
i.e. performing computations with the amplitude of \verb+vnv+ and \verb+vni+
generators set to zero, for a time length of $1 \ s$.

\begin{figure}[!htb]
\includegraphics[width=0.49\columnwidth]{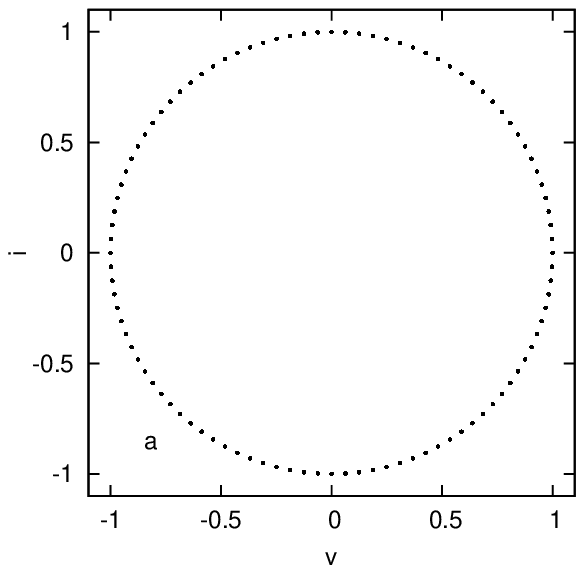}
\includegraphics[width=0.49\columnwidth]{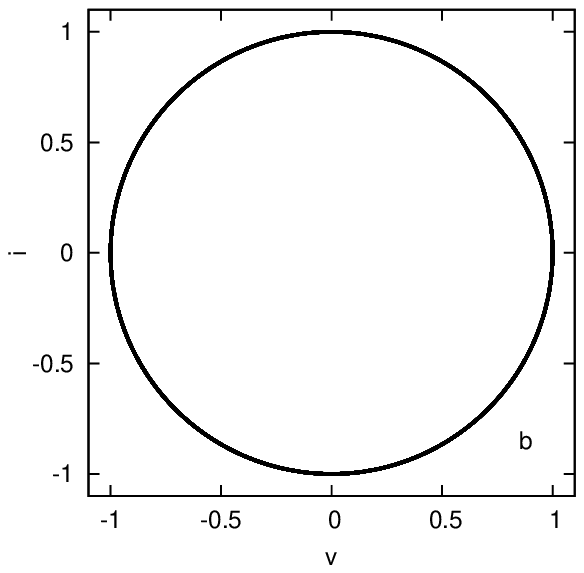}
\caption{
\label{fig-3}
Plot of $i$ vs.\ $v$ without noise (\emph{a}, left)
and with a noise of $0.1 \ V_{rms}$ (\emph{b}, right).
Computational length: $1 \ s$ (1000 cycles).
}
\end{figure}

The results are shown in fig.\ 3\emph{a}, where values in the
data file \verb+data.out+ have been plotted as $i$ vs.\ $v$,
i.e., \verb+i(l1)+ vs. \verb+v(1)+.
The plot covers $1000$ cycles, each cycle containing $100$
equally spaced samples
and confirms phase coherence and amplitude stability
of the computational engine over such a
time length (indeed, a more accurate analysis shows a decay in
amplitude of $0.04\%/s$).
The noise floor due to the overall arithmetic accuracy
is about $-145 \ dBV$ (or $dBA$) in a $1 \ Hz$ bandwidth.

\subsection{Experiment 1: phase fluctuations}

The computations described above
have been repeated activating the noise generators \verb+vnv+
and \verb+vni+
at levels of $0.05$, $0.1$ and $0.2 \ V_{rms}$ over a
time length up to $10 \ s$,
in order to obtain a frequency resolution of $0.1 \ Hz$.
The plot in fig.\ 3\emph{a} changes, in that points do not repeat with the
periodical cadence of $3.6^{\circ}$, rather scatter continuously all
over the circumference.
Neverthless the radius remains unitary, thanks to the behaviour of the
\verb+bvg+ and \verb+big+ generators that suppress the $dP$
component from each couple of $v_n$, $i_n$ values, leaving
only the $dT$ component (fig.\ 3\emph{b}).

\begin{figure}[!htb]
\includegraphics[width=0.85\columnwidth]{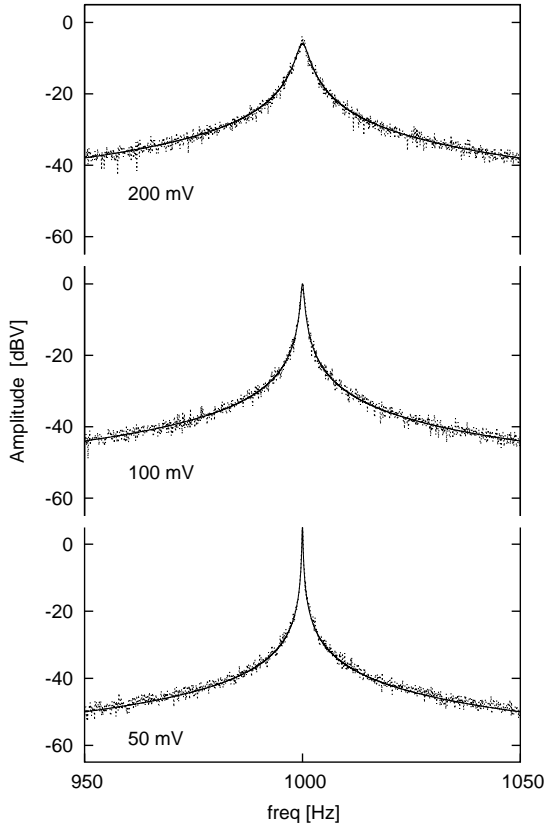}
\caption{
\label{fig:4}
Spectral amplitude in a $1 \ Hz$ bandwidth of the $v$
signal across the $C$ capacitor
in the presence
of the $v_{nt}$ and $i_{nt}$ components only of the $v_n$ and $i_n$ noise
generators.
Noise generators intensity: $50$, $100$ and $200 \ mV$.
Results from the numerical experiment (dots) are plotted
in comparison with Lorentz's function (solid lines).
The $\sigma$ parameter computed from
eq.\ (\ref{eq:Omega-n2}) and (\ref{eq:sigma}) is
$\sigma = 0.0785$, $0.314$ and $1.257 \ Hz$ respectively
from the lower to the upper curve).
No time windowing has been applied with FFT, because the noise
contribution due to truncation is $15$-$20 \ dB$ below the curves
shown in the plot and becomes visible only for noise generators
amplitude of $25 \ mV$ or smaller.
Each curve obtained averaging results from 15 repetitions
of the experiment, with a time length of $10 \ s$.
}
\end{figure}

The spectral amplitude of the generated $v$ signal across capacitor
$C$ is shown in fig.\ 4, together with
the computed Lorentzian distributions, with the $\sigma$ values
obtained from eq.\ (\ref{eq:Omega-n2}) and (\ref{eq:sigma}).
The spectra for current $i$ through inductance $L$ are identical
and are not shown.

The computations have been repeated again using for \verb+vnv+
and \verb+vni+ two generators with a uniform
values distribution instead of a Gaussian one,
with \emph{rms} amplitude $200 \ mV$ (interval $\pm 346.4 \ mV$):
\begin{verbatim}
vnv  3  0  dc 0 trrandom (1 10u 0 346.4m)
vni  4  0  dc 0 trrandom (1 10u 0 346.4m)
\end{verbatim}
and once more with a single gaussian generator, either \verb+vnv+
or \verb+vni+, with an \emph{rms}
amplitude $200 \cdot \sqrt{2} \ mV$.

\begin{figure}
\hbox{
\includegraphics[width=0.51\columnwidth]{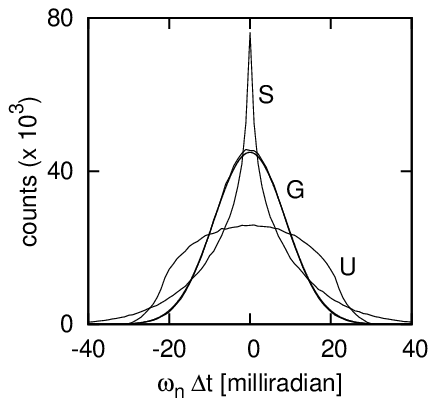}
\includegraphics[width=0.51\columnwidth]{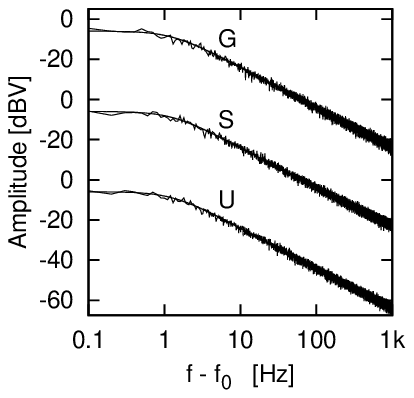}
}
\caption{
\label{fig:5}
Distribution of the $\omega_n \Delta t$ values given by
eq.\ (\ref{eq:omega-n}) (on the left) and Bode plot of the right
branch of the spectral amplitudes (on the right) obtained
with two generators with uniform distribution (U) and
with a single gaussian generator (S).
The third plot (G), for comparison, is made from the data of
fig.\ 4.
The expected analytical gaussian distribution is plotted as well
together with the G curve, on the left, and is practically
undistinguishable from the numerical distribution.
In all cases the total \emph{rms} value of the generators was
\mbox{$200 \cdot \sqrt{2} \ mV$}.}
\end{figure}

The distributions of the $\omega_n \Delta t$ values given by
eq.\ (\ref{eq:omega-n}) for the three cases
are shown in fig.\ 5, on the left.
Only the special case of the first computation exhibits a
Gaussian distribution.
Nevertheless, the other two cases also yield the same identical
Lorentzian spectral amplitudes, as shown in the Bode plots
on the right.
In all cases, the
Gaussian form of the $\varphi_n$ distribution is granted
by effect of the Central Limit Theorem and can be seen
examining the sum of a few consecutive
values of $\omega_n \Delta t$.

\subsection{Experiment 2: amplitude fluctuations
and stabilization}

In order to apply amplitude only fluctuations to the circuit,
without affecting the phase,
generators \verb+bvg+ and \verb+big+ are to be modified
according to eq.\ (\ref{eq:vnp}-\ref{eq:inp}).
In this case $P$ cannot be considered any more a constant, so that
\[
\sin{\varphi} = { i \over {i_0 P}} ~~~~~~~~~~~
\cos{\varphi} = { v \over {v_0 P}}
\]
Thanks to the numerical equivalence of $L$ and $C$ the work can be
done substituting the \verb+bwn+, \verb+bvg+
and \verb+big+ generators in the
Spice list of the first numerical experiment with

\begin{verbatim}
bpq   20  0   v=v(1)*v(1)+i(bvg)*i(bvg)
bpp   21  0   v=(v(4)*v(1)+v(3)*i(bvg))/v(20)
bvg    2  1   v=i(bvg)*v(21)
big    1  0   i=v(1)*v(21)
\end{verbatim}

Here also, generators \verb+bpq+ and \verb+bpp+ are only a convenience,
to avoid the repetition
of a relatively long computation common to both
\verb+bvg+ and \verb+big+.

Fig.\ 6 shows the results obtained with such a machinery with the $P(t)$
plot (right) and the $i$ vs.\ $v$ plot (left).
From the figure it is clear that now the phase proceeds at the constant
angular speed of $3.6^\circ$ per step (i.e., $1000.0 \ Hz$) without
any disturbances, while amplitude fluctuates by effect of energy exchange
with the $v_n$ and $i_n$ generators.
In the long time the oscillation amplitude would fluctuate without
a limit, as usual also for the phase, unless a phase lock mechanism to
some time reference is used.

\begin{figure}[!htb]
\includegraphics[width=1.05\columnwidth]{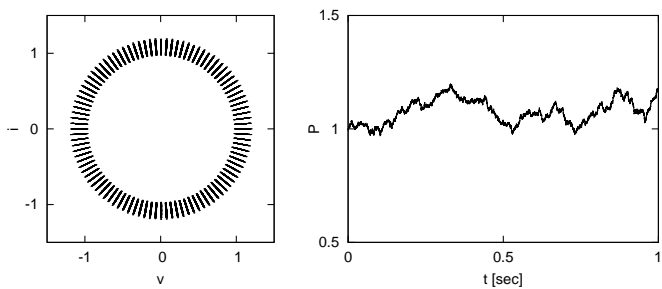}
\caption{
\label{fig:6}
Amplitude fluctuations in the oscillation amplitude during $1 \ s$
with amplitude of {\tt vnv} and {\tt vni} noise generators
set to $0.01 V_{rms}$.
}
\end{figure}

In any case, an amplitude regulation mechanism is necessary
for the reasons already explained.
Even the purely computational approach of Spice applied to an ideal circuit
is not completely free from drift, as quoted in the introduction
to the Numerical Experiments.

The amplitude control expressed by eq.\ (\ref{eq:dPdt})
has been implemented
computing the instantaneous deviation of energy in the circuit
with respect to the reference value $W_0$
\[
\nonumber
\mathcal{E}(t) =
\sqrt{{{1 \over 2} v^2(t) \cdot C + {1 \over 2} i^2(t) \cdot L} \over W_0}
 - 1
\]
and using this quantity as an error signal to actuate the
radial only correction
\begin{equation}
dP = K \mathcal{E}(t) dt
\label{eq:error}
\end{equation}
This can be split into the $dv$ and $di$ components:
\begin{eqnarray}
dv = & v_0 \cos{\varphi} \cdot dP & = v(t) dP \nonumber \\
di = & i_0 \sin{\varphi} \cdot dP & = i(t) dP \nonumber
\end{eqnarray}
Combining these expressions with equation
(\ref{eq:dv-di}) and (\ref{eq:error}), the correction
current and voltage $i_c$ and $v_c$ result:
\begin{eqnarray}
dv = {i_c \over C} dt = v(t) dP ~~ \to ~~ i_c = K C v(t) \mathcal{E}(t)
\label{eq:dvic} \\
di = {v_c \over L} dt = i(t) dP ~~ \to ~~ v_c = K L i(t) \mathcal{E}(t)
\label{eq:divc}
\end{eqnarray}
These corrections
can be applied to the circuit adding their values to generators
{\tt bvg} and {\tt big}, e.g. through generator {\tt bpp},
modifying its definition into
\begin{verbatim}
bpp  21 0  v=(v(4)*v(1)+v(3)*i(bvg))/v(20)+
+                       (sqrt(v(20))-1)*K
\end{verbatim}
and adding the command ``{\tt .param K=...}'' to set the appropriate value
for the {\tt K} constant.

Computations have been performed with an amplitude of $10 \ mV$ for
the {\tt vnv} and {\tt vni} generators and values of
{\tt K} from $10^{-3}$ to 1.
Results are shown in fig.\ 7, together with the expected resonance curves.
The value of $Q_0$ for each value of $K$ has been computed according
to eq.\ (\ref{eq:K}), taking into account that the $C$ and $L$ factors
in eqs.\ (\ref{eq:dvic}) and (\ref{eq:divc})
actually are contained in the computational value of {\tt K}:
\begin{equation}
Q_0 = {{\omega_0} \over {2 K / C}}
\label{eq:Q0}
\end{equation}
\begin{figure}[!htb]
\includegraphics[width=1.0\columnwidth]{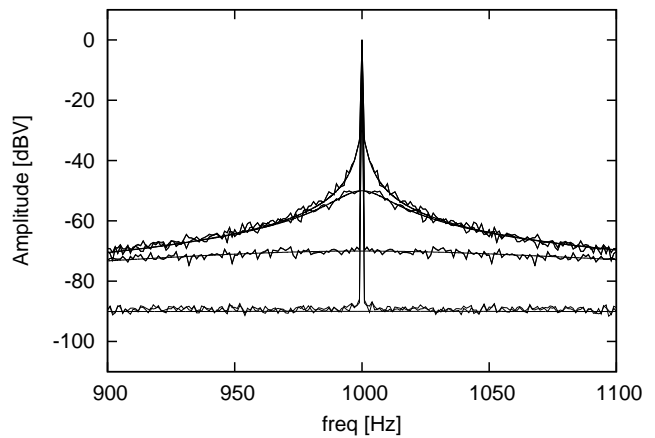}
\caption{
\label{fig:7}
Spectral amplitudes in a $1 \ Hz$ bandwidth of the $v$ signal in the
presence of the $v_{np}$ and $i_{np}$ components only of the $v_n$ and $i_n$
noise generators.
The noisy curves have been obtained with values of the {\tt K} constant
in the amplitude regulation mechanism of $1$, $0.1$, $0.01$ and $0.001$,
from the lower to the upper curve (each one is the average of
25 repetitions of the numerical experiment, with a time length of $1 \ s$).
The smooth lines over the noisy ones are the resonance curves computed
with the $Q_0$ values given by eq.\ (\ref{eq:Q0}).
}
\end{figure}

The frequency spectra clearly show the presence of a really
monochromatic oscillation of amplitude $0 \ dBV$, as set by the initial
conditions, and the response of a resonator with different values
of $Q_0$, excited by the noise signals of $v_n$ and $i_n$.

As already pointed out by most authors, an efficient amplitude
regulation mechanism can reduce the contribution of the amplitude
fluctuations to the overall noise to a level negligible in front of the
phase fluctuations contribution.

However, it is not correct to conclude that half the overall noise
injected by $v_n$ and $i_n$ into the circuit can be suppressed by
the regulation mechanism.
This happens only in this very particular and abstract example 
in which the regulation mechanism uses two synchronized generators
for voltage and current, that act only over the amplitude,
avoiding any interference with the phase.

While it is not impossible by principle to implement practically
such a machinery, it is not that easy.
A possible way, as suggested in
\cite{2003-Ham} is to make the regulating
network, usually behaving as a voltage \emph{or} a current generator,
to interact with the resonator only when the oscillation phase
is at an appropriate point and the correction would require \emph{only}
a voltage \emph{or} a current injection.

Most often it happens that the regulating injected signal contains
a $dT$ component over the $dP$ one and this makes the noise
contribution suppressed as an amplitude fluctuation to appear again
as a phase fluctuation.

Numerical experiments have been performed to verify this point
(results not shown).
Applying both the $dT$ and $dP$ noise components to the resonator
with the amplitude control active (with $K = 10$), noise spectra
identical to fig.\ 4 are obtained.
Thereafter, suppressing the regulating action of one of the
{\tt bvg} or {\tt big} generators, correction is no more only
radial and phase noise spectrum shows the resulting tangential increment.
This is the common behaviour of real oscillators, where the splitting
of the regulation signal
into a $dT$ and a $dP$ components is not usually feasible.

\subsection{Experiment 3: upconversion of low frequency and
$1/f$ noise}

Experiments in this section have been performed in the conditions
of both experiment 1 and 2, i.e. with no amplitude control, applying
only the $dT$ noise component, and with the amplitude control
active (with $K = 10$),
applying both the $dT$ and $dP$ components to the resonator.
This last condition has been implemented using the following
definitions for the {\tt b..} generators:

\begin{verbatim}
bpq  20  0  v=v(1)*v(1)+i(bvg)*i(bvg)
bpp  21  0  v=(sqrt(v(20))-1)*K
bvg   2  1  v=i(bvg)*v(21)+v(3)
big   1  0  i=v(1)*v(21)+v(4)
\end{verbatim}

Identical results have been obtained with both implementations of
the experiment (apart a different level of the noise floor due to the
arithmetics), because in both cases the $dP$ perturbation
component was effectively removed.

\subsection{Experiment 3a}

First, the oscillator behaviour has been studied under the
application of a $10 \ mV$ low frequency sinusoidal signal
with frequency $f_l = 0.25$, $0.5$, $1$, $2$, $4$, $8 \ Hz$ in place of
the {\tt vnv} and {\tt vni} noise generators.
With $f_l = 1 \ Hz$ the generators were
\begin{verbatim}
     vnv  3  0  dc 0 sin(0 10m 1 0 0)
     vni  3  0  dc 0 sin(0 10m 1 0 0)
\end{verbatim}
and the results are shown in fig.\ 8.
 
\begin{figure}[!htb]
\includegraphics[width=1.0\columnwidth]{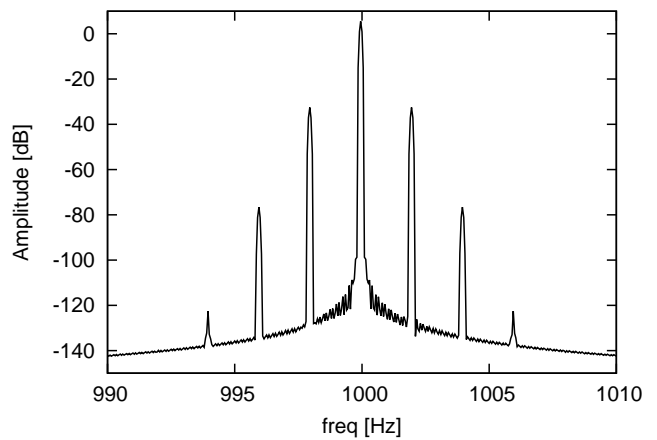}
\caption{
\label{fig-8}
Power spectral density of the $v$ (and $i$) oscillator variable
in the presence of a sinusoidal perturbation at frequency
$1\ Hz$ and amplitude $10\ mV$ for $v_n$ and $i_n$.
The computation has been extended over a time length of $20 \ s$.
A Blackmann-Harris window has been applied with FFT in order to
reduce truncation effects.
This spectrum has been obtained applying ony the $dT$ perturbation
component.
The spectrum obtained the other way, i.e. with the amplitude control
active, shows identical peaks and a noise floor $20 \ dB$ higher.
}
\end{figure}

\subsection{Experiment 3b}

Thereafter, two \emph{coloured} noise sources have been used
for the $v_n$ and $i_n$ generators, obtained summing the usual
white gaussian noise with a $1/f^n$ noise component:

\begin{verbatim}
vnvg  3  30  dc 0 trnoise (0.1 10u 0 0 0 0 0)
vnvf  30  0  dc 0 trnoise (0 10u 1.9 1m 0 0 0)
vnig  4  40  dc 0 trnoise (0.1 10u 0 0 0 0 0)
vnif  40  0  dc 0 trnoise (0 10u 1.9 1m 0 0 0)
\end{verbatim}

\begin{figure}[!htb]
\includegraphics[width=1.0\columnwidth]{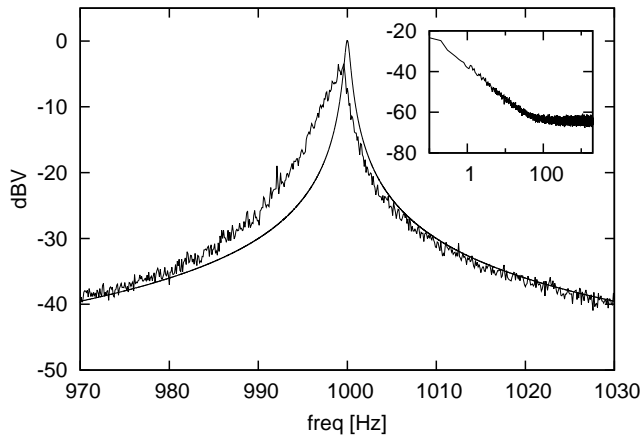}
\caption{
\label{fig-9}
Power spectral density of the $v$ (and $i$) oscillator variable
in the presence of \emph{coloured} noise sources in the
role of the $v_n$ and $i_n$
generators.
The plot inside the inset reports the spectral density of the noise
generators
(averages from 34 computations $10 \ s$ long).
}
\end{figure}

A value close to 2 has been used for the $n$ exponent in the $1/f^n$
distribution to improve the separation among the excess noise region
and the oscillation frequency $f_0$, while keeping an high enough
excess noise amplitude.

Results from this computations are shown in fig.\ 9, together
with the Lorentzian curve associated with the white
gaussian-only noise source.

An analysis of results in fig.\ 8 and 9 will be given in the
Discussion.

\section{Discussion}

Results shown in fig.\ 4 and 7 are well known.
Fig.\ 7 shows that the fluctuations in
the oscillation amplitude can be deeply reduced, to the level of
being negligible, by a suitable circuit regulation mechanism.
When the regulation is tight enough ($K \ge 1$) only a flat
noise floor is left, whose level decreases by $20 \ dB$ for
each tenfold increase in $K$.

The phase component of the noise induced signal fluctuations,
instead, in the case of white noise sources, impresses to the
generated signal spectral distribution the shape of a Lorentzian
curve (fig.\ 4).

The most relevant and new aspect of this part of the work is
the neat and direct mathematical path that starts from the
splitting of the noise into the $dP$ and $dT$ components and
leads to equations (\ref{eq:lorentzian}), (\ref{eq:sigma}) and
(\ref{eq:K}).
These express both the functional form of the noise distribution
and the values of the constants it depends on, directly expressed
in terms of the unique two relevant circuit characteristics:
the energy $W_0$ of the oscillation in the resonator and the rate of
energy exchange (\emph{power}) of the noise source(s)
$I^2_n C + V^2_n L$.

Remaining with the simple case considered in the first numerical
experiment, in which
$${I_n \over {v_0 C}} = {V_n \over {i_0 L}}$$
eq.\ (\ref{eq:Omega-n2}) becomes, e.g., 
\begin{equation}
\Omega_n^2 = {{I^2_n / C} \over {2 W_0}}
\end{equation}
Substituting this last expression into eq.\ \ref{eq:sigma},
\begin{equation}
\sigma =
{1 \over {16 \pi}} \cdot {{I^2_n / C} \over W_0}
\label{eq:sigma-disc}
\end{equation}

If the noise sources were only due
to the losses in the resonator by
effect of the parallel conductance $G$ (or serial resistance $R$, or both),
one should have
$ I_n^2 = 4 K T G $ and, by common formulas,
$Q_0 = \omega_0 C / G$.

Combining this expressions into eq.\ (\ref{eq:sigma-disc}):
\begin{equation}
\sigma =
{1 \over {2 \pi}} {{K T} \over {v_0^2 C}} {{\omega_0} \over Q_0}
\label{eq:sigma-Q0}
\end{equation}

This result is identical to analogous results reported in the literature,
e.g., eq.\ (25) in \cite{2003-Ham} and eq.\ (70) in \cite{1960-Edson};
but in the form of eq.\ (\ref{eq:sigma-disc}) it is more pregnant and
far-reaching.
In a more general case, it is possible through eq.\ (\ref{eq:omega-n})
to sum up into the $i_n$ and $v_n$ generators all noise
contributions that exist in a circuit, then compute
$\Omega_n$ and the resulting line-width $\sigma$.
From the details of the amplitude regulation mechanism, through
eq.\ (\ref{eq:dP}-\ref{eq:inp}), it is possible to compute how
much of the amplitude noise component is converted into phase
noise, to be added to the native phase noise component.

It is worth noting that
the figure of merit $Q_0$ of the resonator does not appear
in eq.\ (\ref{eq:sigma-disc}), while
it appears when passing to eq.\ (\ref{eq:sigma-Q0}).
It has always been believed, and verified experimentally, that an
high value for $Q_0$ is beneficial to reduce phase noise.
Equations (\ref{eq:sigma-disc}) and (\ref{eq:sigma-Q0}) tell
this is true, but only when the losses in the resonator are the
main contribution as noise sources; otherwise, in passing
from eq.\ (\ref{eq:sigma-disc}) to (\ref{eq:sigma-Q0})
$I_n^2$ cannot be replaced by
$4 K T G = 4 K T \omega_0 C / Q_0$, but should become
\begin{equation}
I_n^2 = I_{n0}^2 +
{{4 K T \omega_0 C} \over {Q_0}}
\end{equation}
where $I_{n0}^2$ stands for all other sources of noise
current different from the parallel $G$ of the resonator.
This equation expresses the same concept as the $Q_{loaded}$ in
\cite{2003-Ham}.
Also in \cite{2000-Lee} it had been observed that an increase in $Q_0$ by any
means, that also increase the noise, is vain.

As already observed, what is truly relevant for the spectral purity is the
noise \emph{power} to signal energy ratio as expressed by
eq.\ (\ref{eq:sigma-disc}).

\bigskip

About the third experiment,
the application of a small low frequency perturbation to the circuit,
as made in Experiment $3a$,
reveals the deeply non-linear behaviour induced by any efficient
amplitude regulation mechanism.
Either the radial component $dP$ of the perturbation be removed
\emph{a priori}, as in Experiment 1, or be suppressed by the amplitude
regulation, the identical result is the appearance of the same side-band
components around the frequency of oscillation.

The very interesting aspect to be observed
is that this is not a simple effect
of amplitude modulation (AM):
the plot in fig.\ 9 clearly shows that lateral components are at
frequency $1000 \pm 2 \ Hz$, $\pm 4 \ Hz$ and $\pm 6 \ Hz$, hence at an
offset frequency twice the $1 \ Hz$ frequency of the perturbation
signal (and higher order harmonics).
True AM sidebands can be obtained, instead, adding the modulating signal
to the expression {\tt sqrt(20)-1} in the {\tt bpp} generator
definition (results not shown).

A close analysis of the data in fig.\ 9 clearly shows that the
central peak is no more exactly at $1000 \ Hz$, as it
sistematically was in all previous computations;
rather, it is slightly moved to the left by a small amount
that can be appreciated also in the plot.
As a fact, the {\tt data.out} file obtained from this experiment does
contain a signal with average frequency of $999.95 \ Hz$ instead of
1000.00.
This offset in the oscillator frequency depends upon the amplitude
of the perturbing signal but does not depend upon its frequency.
All frequencies from $0.25$ to $8 \ Hz$ gave exactly the same offset;
varying the amplitude in the $1 - 10 \ mV$ range a dependance
of the offset with the square of the amplitude was observed.

Reducing the value of the $K$ constant while performing the
third experiment in the condition of Experiment 2, the sidebands
and the frequency offset progressively reduce and vanish with
$K \to 0$.

This makes sense, because with $K \to 0$ every nonlinear behaviour
is removed from the circuit and only a linear resonator is left,
excited by a voltage/current at frequency $f_l$.
In the signal spectrum only $f_0$ and $f_l$ can be found, with no
mixing effect at all.

The frequency doubling cannot be an
artifact due to the particular amplitude regulation algorithm
adopted; it appears identical with no regulation mechanism at all,
only by effect of noise splitting and $dP$ component suppression.

These findings are to be kept in mind when observing the behaviour
of an oscillator in the presence of a \emph{coloured} noise source,
as it is the case for $1/f$ noise components.
Results in fig.\ 9 (the noisy line) show the behaviour in such a condition,
compared to the white noise behaviour of fig.\ 4.
It is not a simple process of upconversion and appearance of an upper and
lower band on side of $f_0$; rather, it is a deformation and a shift
to the left of the distribution and its peak due to a process of frequency
pulling as described in \cite{2004-Razavi,2010-Mirzaei}.

In view of these results, it is doubtful that this process can be
considered the source (or the unique source) of $1/f$ components
observed in oscillators on side of the carrier.
Other effects should be considered;
for example, the direct modulation of the generated signal in passing
through a component with equilibrium value affected by $1/f$
fluctuations \cite{1976-Voss}.
But this is argument for a new investigation and is out of the
purposes of this paper.

\section{Acknowledgments}

The Florence section of INFN (Istituto Nazionale di Fisica Nucleare) and
Prof. Maurizio Bini are gratefully acknowledged for granting
the access to the 64-core computing machine used in most of the
described computations.

\end{document}